# An explanation for a universality of transition temperatures in families of copper-oxide superconductors

**Sudip Chakravarty[1], Hae-Young Kee[2] & Klaus Völker[2]**

*[1]Department of Physics and Astronomy, University of California Los Angeles, Los Angeles, California 90095-1547, USA*

*[2]Department of Physics, University of Toronto, Ontario M5S 1A7, Canada*

**A remarkable mystery of the copper oxide high-transition-temperature ($T_c$) superconductors is the dependence of $T_c$ on the number of $CuO_2$ layers, $n$, in the unit cell of a crystal. In a given family of these superconductors, $T_c$ rises with the number of layers, reaching a peak at $n$=3, and then declines[1]: the result is a bell-shaped curve. Despite the ubiquity of this phenomenon, it is still poorly understood and attention has instead been mainly focused on the properties of a single $CuO_2$ plane. Here we show that the quantum tunnelling of Cooper pairs between the layers[2] simply and naturally explains the experimental results, when combined with the recently quantified charge imbalance of the layers[3] and the latest notion of a competing order[4–9] nucleated by this charge imbalance that suppresses superconductivity. We calculate the bell-shaped curve and show that, if materials can be engineered so as to minimize the charge imbalance as $n$ increases, $T_c$ can be raised further.**

The phase diagrams of the high-$T_c$ superconductors show two striking universalities. The first is the dependence of $T_c$ on the number of $CuO_2$ layers, $n$, within a homologous series, as shown in Fig. 1. The second is the superconducting dome: as a function of doping, $x$ (charge carriers added to the conducting $CuO_2$ planes), $T_c$ is a bell-shaped curve rising at about $x \approx 0.05$ and dropping to zero at $x \approx 0.3$; see Fig. 2 for a similar behaviour of the superconducting order parameter.

It is also evident that these superconductors are rife with competing orders, of which the most recent experimental demonstration is the measurement of the Hall number in the 60 T magnetic field[10]. A specific suggestion has been that the demise of superconductivity, and hence the universal origin of the superconducting dome as a function of $x$, is due to a new competing order called the $d$-density wave (DDW)[8], where a particle and a hole are bound in the angular momentum channel of $l$=2 ($d$-wave). The second universal feature was also recognized previously, and interlayer tunnelling theory was developed to explain it[2]. Unfortunately, this theory leads to a $T_c$ that saturates as a function of $n$[11], which is not the observed bell-shaped curve shown in Fig. 1. In addition, a particularly strong conjecture[12] that the entire pairing mechanism was due to interlayer tunneling was later called into question by experiments[13].

In contrast to ref. 12, we argue that the tunnelling between the layers must be regarded as a mechanism by which pairing is enhanced[14,15] and should not be construed as the sole reason for high $T_c$, especially for single-layer materials in which the effect of tunnelling is negligible. To take advantage of tunnelling between the close pairs of $CuO_2$ planes within a unit cell, it is necessary that a low-energy electron in the normal state is forbidden to tunnel coherently perpendicular to the planes[2]. In the superconducting state this kinetic energy is recovered, resulting in enhanced pairing. The frustrated kinetic energy in the normal state may either be due to a non-Fermi-liquid nature of this state, in which an electron breaks up into more fundamental constituents[2], or due to the pseudogap that is present over much of the phase diagram[16]. From a simple renormalization group argument, in the pseudogap state, the single-particle tunnelling is irrelevant at low energies below the gap, and it should simply drop out of all macroscopic considerations[17]. In contrast, pair tunnelling, in which a Cooper pair of electrons tunnels together, is a coherent zero-energy process and must have macroscopic consequence. For high-$T_c$ superconductors, the tunnelling matrix element in the momentum space is peaked where the superconducting gap is large[2], and therefore the existence of nodes in the gap cannot invalidate this argument. The importance of pair tunnelling is also crucial in stripe theories of superconductivity[18].

We shall not dwell further on microscopic considerations, because the important aspect of the phase diagram can be understood using a suitable free-energy functional. Imagine that two-dimensional planes, indexed by $j$, are stacked to form unit cells consisting of $n$ layers of a three-dimensional superconductor such that the mean-field free-energy function of the complex order parameters $\psi_j$ in a unit cell is ($A$ is the area of the two-dimensional plane):

$$F_s = A \sum_j \left[ \alpha' |\psi_j|^2 + \lambda' |\psi_j|^4 - \rho_c \left( \psi_j \psi_{j+1}^* + \text{c.c.} \right) \right] \tag{1}$$

where $\alpha'$, $\lambda'$, and $\rho_c$ are parameters that may in principle depend on the layer index, and the gradient terms are left out in mean field theory; 'c.c.' stands for complex conjugation. At temperature $T = 0$, $F_s$ is the ground-state energy. On symmetry grounds, there are no distinctions between the following two choices of the coupling between the layers: $-\left( \psi_j \psi_{j+1}^* + \text{c.c.} \right)$ and $\left| \psi_j - \psi_{j+1} \right|^2$. They are simply related by a shift of the quadratic term.

Our tacit choice results in an enhancement of pairing, while the choice $\left|\psi_j - \psi_{j+1}\right|^2$ does not, because its minimum value is zero[19]: the crux, of course, is to enhance the mean-field $T_c$, as fluctuations can only reduce it. Equation (1) results in a $T_c$ that initially rises with $n$ and then saturates[11]. Although the rise can be significant, the crucial subsequent drop and the maxima at $n = 3$ are not captured.

We now include the free energy of the competing order parameter that is ultimately responsible for the downturn of $T_c$ with $n$. For concreteness, consider a model where the competing order is DDW, but any other suitable order[20] may also serve the same purpose. For the DDW order parameter $\phi_j$ that breaks the time reversal symmetry and couples ($g$ is the coupling) to the superconducting order parameter $\psi_j$, the free energy is:

$$F_c = A \sum_j \left[ \alpha \phi_j^{\,2} + \lambda \phi_j^{\,4} + g \left|\psi_j\right|^2 \phi_j^{\,2} \right] \qquad (2)$$

The parameters $\alpha$, $\lambda$ and $g$ may depend on the layer index. The coupling between $\psi$ and $\phi$ is a fourth-order invariant, and therefore, to be consistent, the individual fourth-order terms in equations (1) and (2) must not be neglected. We have ignored the negligibly small interlayer coupling of the DDW order parameter $\phi_j$ (ref. 21).

Another important ingredient in our theory is recent nuclear magnetic resonance (NMR) measurements of the Hg-series[3,22,23]. Using an empirical relation between the spin part of the $^{63}$Cu Knight shift and the hole concentration for materials with $n$=1 and $n$=2, the doping of the individual planes up to $n$=5 was deduced. It was found that in a unit cell the outer layers tend to get overdoped, while the inner layers tend to get underdoped, consistent with simple but robust electrostatic Madelung energy considerations[3].

The complete Ginzburg–Landau free energy at $T = 0$ of a $n$-layer system, including the doping imbalance, is given by $F = F_{\mathrm{s}} + F_{\mathrm{c}}$, where:

$$F = A \sum_j \left[ \alpha'(x_j) |\psi_j|^2 + \lambda' |\psi_j|^4 - \rho_{\mathrm{c}} \left( \psi_j \psi_{j+1}^* + \mathrm{c.c.} \right) + \alpha(x_j) \phi_j^2 + \lambda \phi_j^4 + g |\psi_j|^2 \phi_j^2 \right] \quad (3)$$

and $\alpha'(x_j)$ and $\alpha(x_j)$ are functions of the doping $x_j$ for each layer; all other parameters are assumed to be constants. For a single-layer system, there is only one term in the sum, $x_j \equiv x$, $\rho_{\mathrm{c}} \equiv 0$, and the minimum of the free energy is easily found analytically. The free-energy functional in equation (3) gives the right shape and the magnitude of the superconducting dome for a generic single-layer material, as shown in Fig. 2.

In the general case we have to resort to numerical minimization. We use a downhill simplex method combined with simulated annealing to determine the individual order parameters for each layer in a unit cell, with an open boundary condition. As indicators for the superconducting transition temperature, we have examined the following three criteria: $T_{\mathrm{c}} \propto \psi_{\mathrm{avg}} = \frac{1}{n} \sum_1^n |\psi_j|$, $T_{\mathrm{c}} \propto \psi_{\mathrm{r.m.s.}} = \left( \frac{1}{n} \sum_1^n |\psi_j|^2 \right)^{1/2}$, and $T_{\mathrm{c}} \propto \psi_{\mathrm{max}} = \max |\psi_j|$. The choice between these three indicators is very complex, involving many details related to the proximity effect. In a nutshell, $\psi_{\mathrm{avg}}$ is a better indicator when the superconducting coherence length in a direction perpendicular to the planes is relatively large, whereas $\psi_{\mathrm{max}}$ is more appropriate in the opposite limit when this length is very short. Fortunately, as we shall see, the robust features of the phase diagram are independent of the choice. We have refrained, however, from using a finite-temperature version of the free-energy function, because it would have introduced additional adjustable parameters. The parameters in equation (3) are strongly constrained by the physics of the single-layer problem. Figure 3 shows the dependence of $T_c$ on the number of layers. The results capture the dependence found in experiments: interlayer coupling enhances the transition temperature up to $n$=3, but the large doping imbalance, combined with the competing order, reduces the optimal $T_c$ beyond three layers.

According to the theory of competing order[8], the total single-particle excitation gap, $E_g$, is not the superconducting gap alone, but a function of both the superconducting gap and the gap due to the hidden order, the DDW gap for instance. Whereas the superconducting gap is controlled by the order parameter, $\psi$, the competing gap is controlled by $\phi$. Thus, $E_g$ can be finite even when $\psi$ is zero. This is different from those theories where the total gap is the local superconducting gap alone, and the decrease in $T_c$ with $x$ is due to phase fluctuations[24]. Although phase fluctuations do play a role in determining $T_c$, in our view it is the competition of the two order parameters that plays the dominant role in determining the superconducting dome.

In principle, we can further adjust parameters to fit data better, but it would not be very meaningful for two reasons. First, the actual enhancement of $T_c$ for different homologous series differs in magnitude. Second, we have only calculated the mean-field order parameters at $T$=0. Fluctuation effects will surely be important for the actual transition temperatures. Fluctuations will depress the $T_c$ for $n$=1 more than the $T_c$ for $n$=3. Thus, the overall increase in $T_c$ will be in better agreement with experiments. We note that conventional Josephson coupling between the layers, $\left| \psi_j - \psi_{j+1} \right|^2$, can suppress fluctuations and raise the actual transition temperature closer to the mean-field value as the layers are

coupled[19]. This mechanism was indeed explored[25], but the enhancement of $T_c$ was too small to explain the striking data shown in Fig. (1). Moreover, a mechanism to explain the bell-shaped profile of $T_c$ as a function of $n$ was missing in this calculation.

According to our theory, to increase $T_c$ further, it would be necessary to dope the system in such a way that the layers do not develop a charge imbalance and nucleate competing order. Another consequence would be that site-specific NMR relaxation rates should show a pseudogap in the inner layers, but not in the outer layers, owing to the charge imbalance, which is in agreement with recent experiments[22,23]. This is because the suppression of the superconducting order parameter in the inner layers will enhance the pseudogap. The single-particle excitation spectra as observed in angle resolved photoemission spectroscopy (ARPES) of multilayer copper oxides should be sensitive to the doping imbalance of the layers; there is already some indication of this in experiments[26,27] involving the triple-layer material Bi2223. Although bilayer splitting is observed in optimally doped Bi2212, trilayer and higher-multilayer splittings will be increasingly difficult to observe, because of the induced pseudogap of the inner layers.

We also predict that the maximum superconducting gap measured in ARPES will be a bell-shaped curve as a function of $n$ with a maximum at $n = 3$. It would be worth investigating whether the recently developed Fourier-transform scanning tunnelling spectroscopy (FT-STS)[28] could provide layer-specific information. The change in the spectra with increasing $n$ should be detectable, as the tunnelling rate falls off exponentially with the distance. For this, it would be interesting to consider an underdoped sample. As $n$ increases, the spectra, which would be dominated by the outer layer, will change because of its increased doping.

**Acknowledgements** We thank the Aspen Centre for Physics, where this collaboration was initiated, and also N. P. Armitage and J. Hoffman for discussions. We acknowledge support from the US National Science Foundation (S.C.), the Canadian Institute for Advanced Research (H.-Y.K. and K.V.), and the Alfred P. Sloan Foundation (H.-Y.K.).

**Competing interests statement** The authors declare that they have no competing financial interests.

**Correspondence** and requests for materials should be addressed to S.C. (sudip@physics.ucla.edu).

**Figure 1** Transition temperature within a homologous series. A homologous series[29] is a family of compounds having the same charge-reservoir block, but $n CuO_2$-planes in the infinite-layer block, which in turn consists of $(n-1)$ bare cation planes and $n CuO_2$-planes. A good example is the family $HgBa_2Ca_{(n-1)}Cu_nO_{(2n+2+\delta)}$ whose $T_c$ as a function of $n$, optimized with respect to oxygen concentration, is shown. (The figure is adapted from the data in ref. 30). Similar results have been known for some time; see ref. 1.

**Figure 2** The $T$=0 phase diagram of a one-layer copper oxide as a function of doping, $x$. To reproduce the well-known superconducting dome of the order parameter of a single-layer copper oxide, we choose $\alpha(x) = 27(x - 0.22)$, $\alpha'(x) = 10(x - 0.3)$, $\lambda = \lambda' = 1$ and $g = 1.2$. These are the same parameters as those in ref. 8. The order parameter $\psi$ is the superconducting order parameter, and $\phi$ is the competing order parameter.

**Figure 3** The calculated superconducting order parameters at $T$=0 of multilayer copper oxides. For $n \geq 3$, the parameters $\alpha(x_j)$ and $\alpha'(x_j)$ in equation (3) vary with the layers because of the charge imbalance between the inner and the outer layers in a unit cell. Once we know the doping of a layer $x_j$, we use the same parameterizations as in the legend of Fig. 2. Thus, the superconducting dome is entirely determined by the physics of the competing order of a single layer material. In accordance with ref. 3, we assume that an amount of charge $\varepsilon$ is transferred from the ($n$–2) inner layers to the two outer layers, so that the effective doping is $x_{\mathrm{I}} = \left[1 - \varepsilon/(n - 2)\right]x$ on the inner layers, and $x_{\mathrm{O}} = \left[1 + \varepsilon/2\right]x$ on the outer layers. Here $x$ is the average, or nominal, doping per layer. The charge imbalance $\varepsilon$ is extracted from ref. (3), where the ratio $x_{\mathrm{O}}/x_{\mathrm{I}} = R_{\mathrm{h}}$ is given by 1.14 for $n$=3, 1.49 for $n$=4, and 1.64 for $n$=5, with $x \approx 0.2$. Because $\varepsilon = \left[2(n - 2)(R_{\mathrm{h}} - 1)\right]/(n - 2 + 2R_{\mathrm{h}})$, we obtain $\varepsilon =$ 0.085, 0.39 and 0.61 for the 3-, 4- and 5-layer systems, respectively. The interlayer coupling $\rho_{\mathrm{c}}$ is the only remaining free parameter, which we set to be 0.3 to optimize the shape of $T_{\mathrm{c}}$ versus $n$. Here $\psi_{\mathrm{avg}}$ (circles), $\psi_{\mathrm{r.m.s}}$ (triangles) and $\psi_{\mathrm{max}}$ (squares) are shown as a function of the number of layers $n$, as the indicators for the superconducting transition temperature $T_{\mathrm{c}}$. The error bars were supplied by Y. Kitaoka (personal communication).

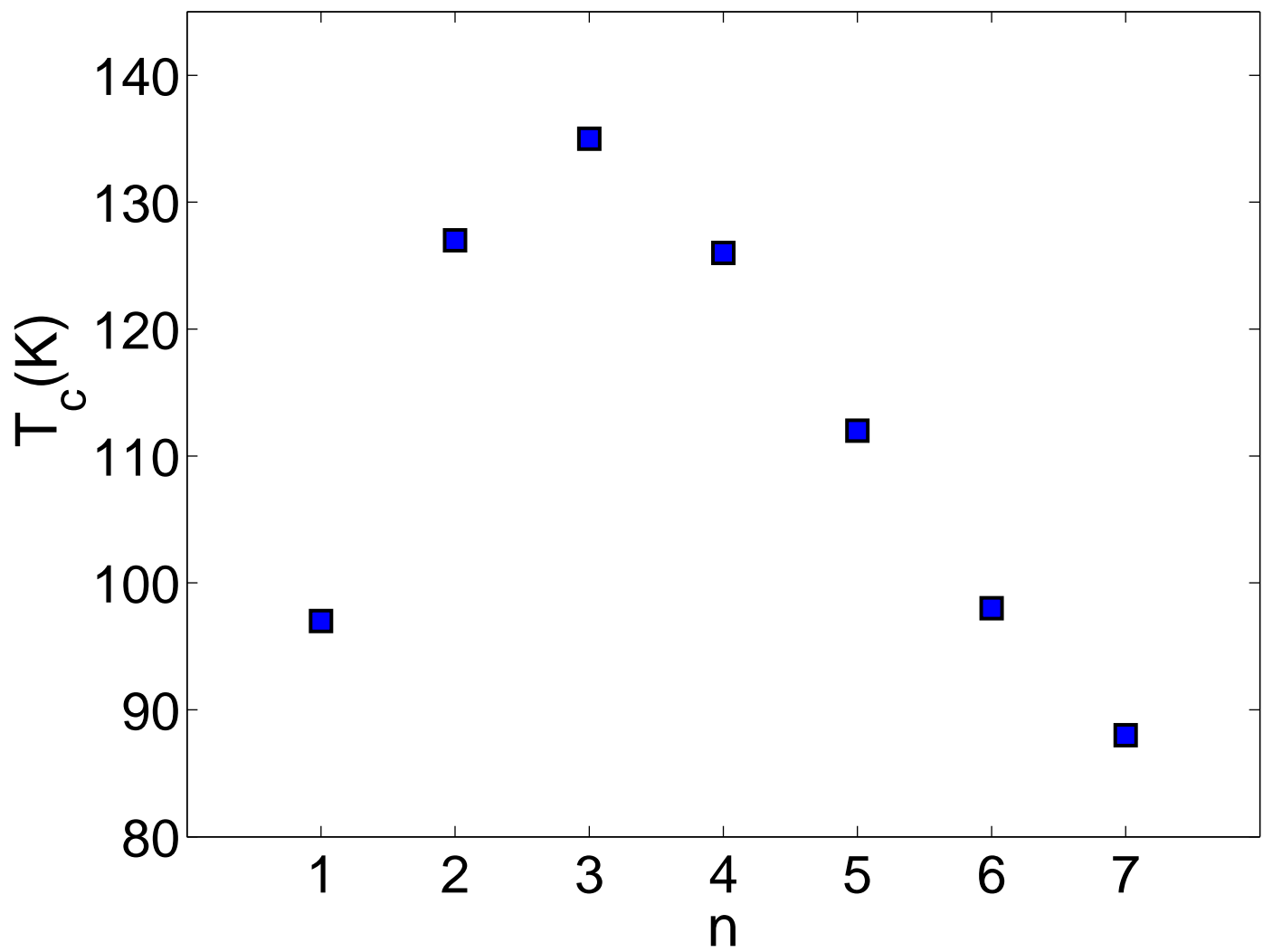

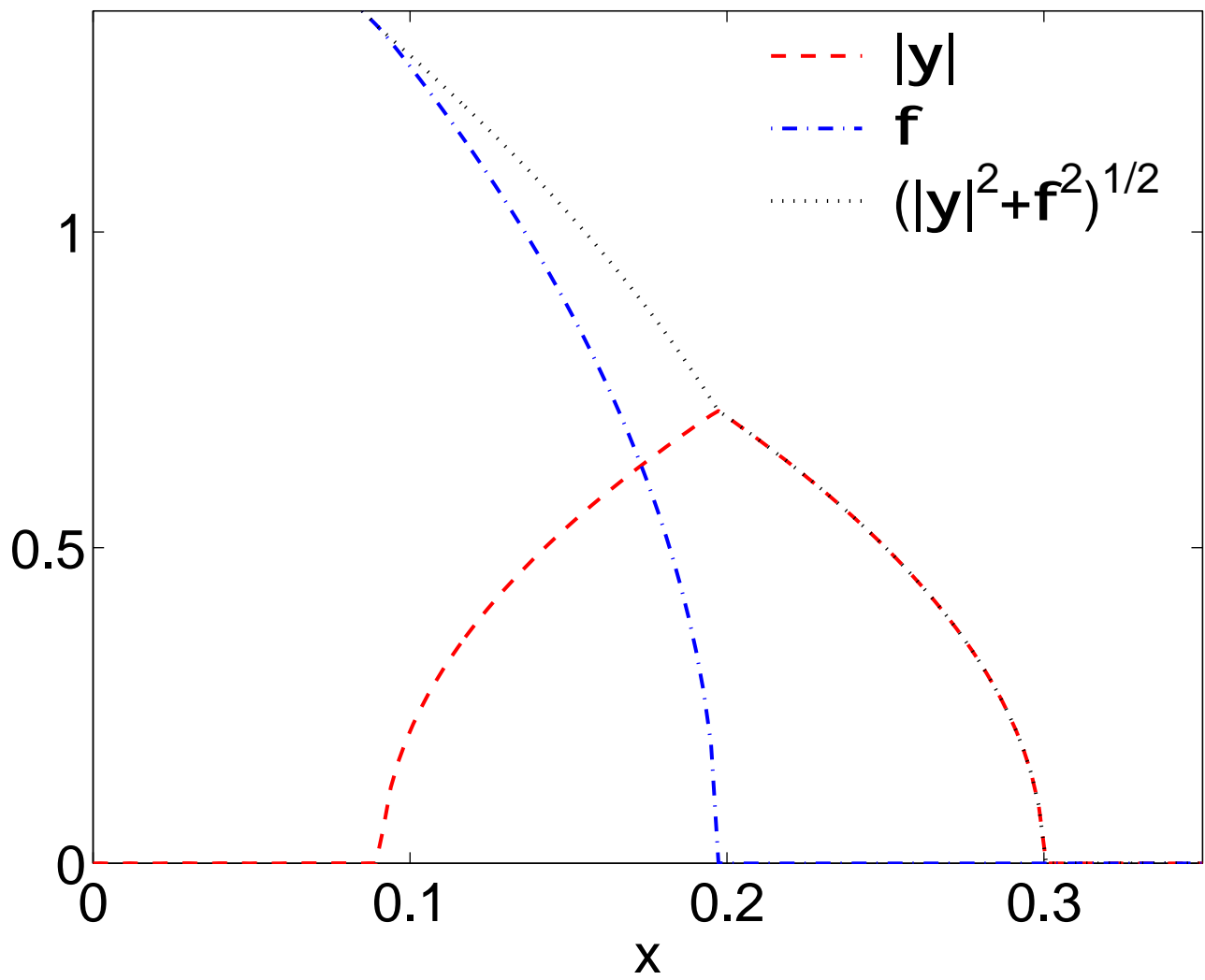

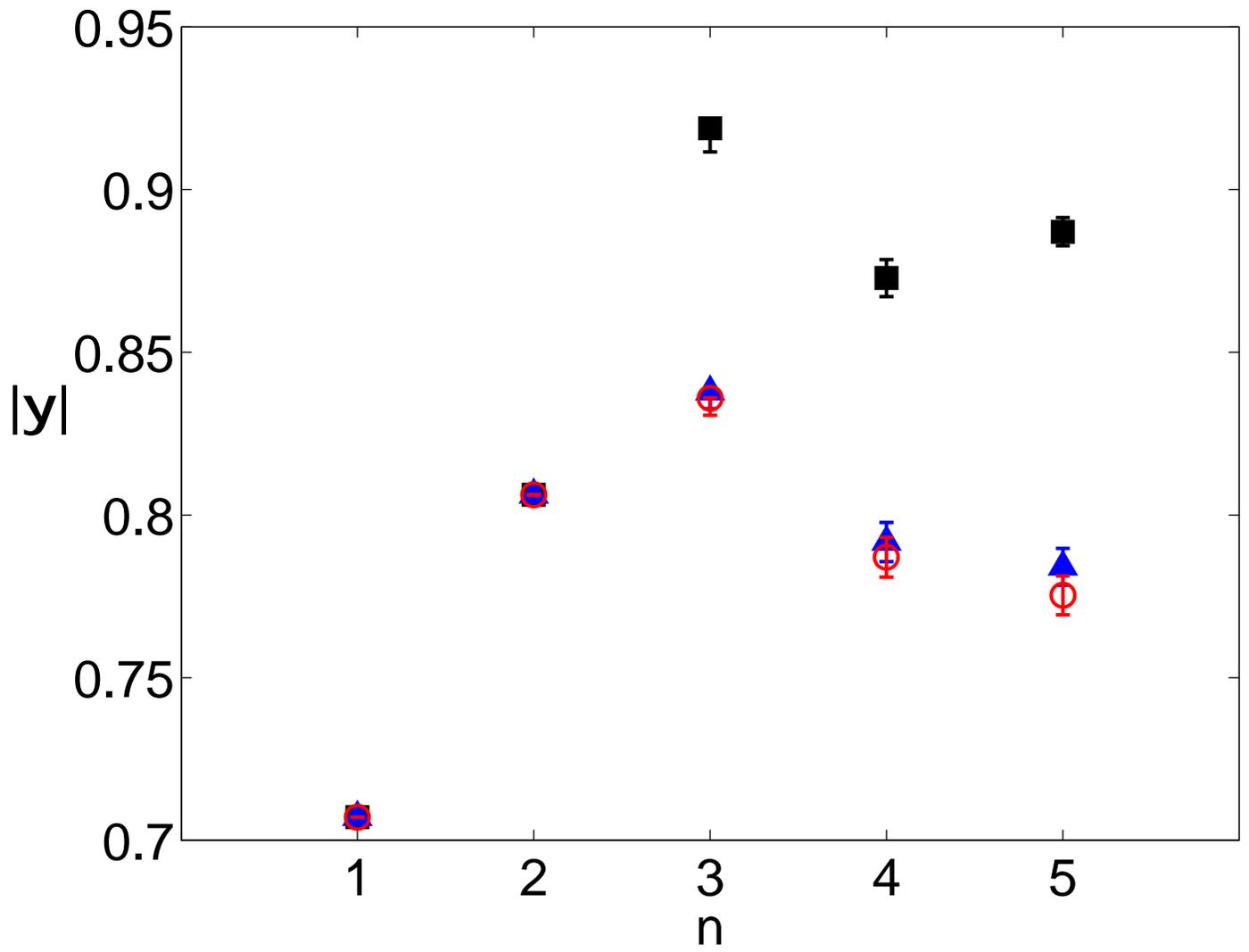